\documentclass[12pt]{article}
\usepackage{graphicx}
\input epsf
\def\hybrid{\topmargin 0pt      \oddsidemargin 0pt
        \headheight 0pt \headsep 0pt
        \voffset=-0.5cm
        \textwidth 6.25in       
        \textheight 9.5in       
        \marginparwidth 0.0in
        \parskip 5pt plus 1pt   \jot = 1.5ex}
\catcode`\@=11
\def\marginnote#1{}

\newcount\hour
\newcount\minute
\newtoks\amorpm
\hour=\time\divide\hour by60
\minute=\time{\multiply\hour by60 \global\advance\minute by-\hour}
\edef\standardtime{{\ifnum\hour<12 \global\amorpm={am}%
        \else\global\amorpm={pm}\advance\hour by-12 \fi
        \ifnum\hour=0 \hour=12 \fi
        \number\hour:\ifnum\minute<10 0\fi\number\minute\the\amorpm}}
\edef\militarytime{\number\hour:\ifnum\minute<10 0\fi\number\minute}

\def\draftlabel#1{{\@bsphack\if@filesw {\let\thepage\relax
   \xdef\@gtempa{\write\@auxout{\string
      \newlabel{#1}{{\@currentlabel}{\thepage}}}}}\@gtempa
   \if@nobreak \ifvmode\nobreak\fi\fi\fi\@esphack}
        \gdef\@eqnlabel{#1}}
\def\@eqnlabel{}
\def\@vacuum{}
\def\draftmarginnote#1{\marginpar{\raggedright\scriptsize\tt#1}}
\def\draftlabel#1{{\@bsphack\if@filesw {\let\thepage\relax
   \xdef\@gtempa{\write\@auxout{\string
      \newlabel{#1}{{\@currentlabel}{\thepage}}}}}\@gtempa
   \if@nobreak \ifvmode\nobreak\fi\fi\fi\@esphack}
        \gdef\@eqnlabel{#1}}
\def\@eqnlabel{}
\def\@vacuum{}
\def\draftmarginnote#1{\marginpar{\raggedright\scriptsize\tt#1}}

\def\draft{\oddsidemargin -.5truein
        \def\@oddfoot{\sl preliminary draft \hfil
        \rm\thepage\hfil\sl\today\quad\militarytime}
        \let\@evenfoot\@oddfoot \overfullrule 3pt
        \let\label=\draftlabel
        \let\marginnote=\draftmarginnote
   \def\@eqnnum{(\theequation)\rlap{\kern\marginparsep\tt\@eqnlabel}%
\global\let\@eqnlabel\@vacuum}  }

\def\numberbysection{\@addtoreset{equation}{section}
        \def\theequation{\thesection.\arabic{equation}}}

\def\underline#1{\relax\ifmmode\@@underline#1\else
        $\@@underline{\hbox{#1}}$\relax\fi}

\def\titlepage{\@restonecolfalse\if@twocolumn\@restonecoltrue\onecolumn
     \else \newpage \fi \thispagestyle{empty}\c@page\z@
        \def\thefootnote{\fnsymbol{footnote}} }

\def\endtitlepage{\if@restonecol\twocolumn \else  \fi
        \def\thefootnote{\arabic{footnote}}
        \setcounter{footnote}{0}}  
\relax

\numberbysection
\hybrid

\newfont{\Bbb}{msbm10 scaled 1\@ptsize00}
\newcommand{\CC}{\mbox{\Bbb C}}

\newfont{\Bbbb}{msbm7 scaled 1\@ptsize00}
\newcommand{\cc}{\raise-1pt\hbox{$\mbox{\Bbbb C}$}}
\newcommand{\zz}{\raise-1pt\hbox{$\mbox{\Bbbb Z}$}}

\def\beq{\begin{equation}}
\def\eeq{\end{equation}}
\def\p{\partial}

\def\DD{{\sf D}}

\def\unitary{{\cal U}}

\def\normal{{\cal N}}
\def\complex{{\cal C}}

\def\lbracket{\left <}
\def\rbracket{\right >}
\def\vphcl{\varphi _{0}}
\def\rhocl{\rho _{0}}

\begin{document}
\begin{titlepage}

\title{Random matrices and Laplacian growth}

\author{A.~Zabrodin
\thanks{Institute of Biochemical Physics,
4 Kosygina st., 119334, Moscow, Russia and ITEP, 25
B.Cheremushkinskaya, 117218, Moscow, Russia}}

\date{July 2009}
\maketitle

\begin{abstract}
The theory of random matrices with eigenvalues distributed
in the complex plane and more general ``$\beta$-ensembles"
(logarithmic gases in 2D) is reviewed.
The distribution and correlations of the eigenvalues
are investigated in the large $N$ limit.
It is shown that in this limit the model
is mathematically equivalent to a class of
diffusion-controlled growth
models for viscous flows in the Hele-Shaw cell and
other growth processes of Laplacian type.
The analytical methods used involve
the technique of boundary value problems
in two dimensions and elements of the potential theory.

\end{abstract}

\vfill

\end{titlepage}

\section{Introduction}

Applications of random matrices
in physics (and mathematics)
are known to range from energy levels
statistics in nuclei to number theory and from quantum
chaos to string theory.
Most extensively employed
and best-understood are ensembles of hermitian
or unitary matrices (see e.g.
\cite{Mehta}-\cite{Morozov}). Their eigenvalues are confined
either to the real axis or to the unit circle.
In this paper we consider more general
classes of random matrices, with no a priori restrictions
to their eigenvalues.
Such models are as yet less well understood but
they are equally interesting and meaningful from
both mathematical and physical points of view.
(A list of the relevant
physical problems and corresponding references
can be found in, e.g., \cite{list1}.)

The progenitor of ensembles of matrices with general
complex eigenvalues is the statistical model of
complex matrices with the Gaussian
weight introduced by Ginibre \cite{Ginibre} in 1965.
The partition function of this model is
\beq\label{intro1}
Z_N = \int [D\Phi ] \exp \left ( -\, \frac{N}{t}
\mbox{tr}\, \Phi^{\dag} \Phi \right ).
\eeq
Here $[D \Phi ] = \prod_{ij}d(\Re \, \Phi_{ij})
d(\Im \, \Phi_{ij})$ is the standard volume element
in the space of $N\times N$ matrices with complex
entries $\Phi_{ij}$ and $t$ is a (real positive)
parameter.
Along with the Ginibre
ensemble and its generalizations
we also consider ensembles of normal
matrices \cite{normal}, i.e., such that $\Phi$ commutes with
its hermitian conjugate $\Phi^{\dag}$.

Since one is primarily interested in statistics of eigenvalues,
it is natural to express the probability density
in terms of complex
eigenvalues $z_j =x_j + i y_j$
of the matrix $\Phi$.
It appears that the volume element can be represented as
\beq\label{intro2}
[D \Phi ] \propto \prod_{i<j}
\left |z_i - z_j \right |^{2}\,
\prod_i d^2 z_i.
\eeq
If the statistical weight
depends on the eigenvalues only, as it is usually
assumed, the other parameters of the matrix
(often referred to as
``angular variables") are irrelevant
and can be integrated out giving an overall normalization
factor. In this case the original matrix problem
reduces to statistical mechanics of $N$ particles
with complex coordinates $z_j$ in the plane.
Specifically, the factor
$\prod_{i<j}
\left |z_i - z_j \right |^{2}$,
being equal to the exponentiated Coulomb energy
in two dimensions, means an
effective ``repelling" of eigenvalues.
This remark leads to the Dyson logarithmic
gas interpretation \cite{Dyson}, which treats the matrix
ensemble as
a two-dimensional ``plasma" of eigenvalues in a
background field and prompts to introduce
more general ``$\beta$-ensembles" with the statistical
weight proportional to $\prod_{i<j}
\left |z_i - z_j \right |^{2\beta}$.

It is also natural to consider matrix ensembles with
statistical weights of a general form,
$\displaystyle{\exp \left ( -\, \frac{N}{t}
\mbox{tr}\, W(\Phi^{\dag}, \Phi )\right )}$,
with a background potential $W$.
An important observation made in \cite{KKMWZ}
and developed in subsequent works \cite{WZnormal,Z03,TBAZW05,Z06}
is that evolution of an averaged
spectrum of such matrices as a function
of $t$, as $N \to \infty$, serves as a simulation of
Laplacian growth of water droplets
in the Hele-Shaw cell (for different physical and
mathematical aspects of the latter see \cite{RMP,book,list}).
To be more precise, in the limit
$N\to \infty$,
under some weak assumptions about the
statistical weight,
the eigenvalues are confined, with probability 1, to a
compact domain in the complex plane. One can ask how
its shape depends on the parameter $t$.
The answer is: this dependence
is exactly the Laplacian growth of the domain with
zero surface tension.
Namely,
the edge of the support moves along gradient of a
scalar harmonic field
in its exterior, with the velocity being
proportional to the
absolute value of the gradient. The general solution
can be expressed in term of the exterior Dirichlet boundary value
problem.

This fact allows one to treat the model of normal or complex
random matrices as a growth problem \cite{TBAZW05}. The advantage of this
viewpoint is two-fold. First, the hydrodynamic interpretation
makes some of the large $N$ matrix model results
more illuminating
and intuitively accessible. Second and most important,
the matrix model perspective
may help to suggest new approaches to the
long-standing growth problems. In this respect,
of special interest is the identification
of finite time singularities in some exact solutions
to the Hele-Shaw flows with critical points of the normal
and complex matrix models.

\section{Random matrices
with complex eigenvalues}

We consider square random matrices $\Phi$
of size $N$
with complex entries $\Phi_{ij}$.
The probability density is assumed to be
of the form $P(\Phi )\propto e^{\frac{1}{\hbar}{\rm tr} W(\Phi )}$, where
the function $W(\Phi )$ (often called the potential of
the matrix model) is a matrix-valued
function of $\Phi$ and $\Phi^{\dag}$ such that
$(W(\Phi ))^{\dag} =W(\Phi )$ and $\hbar$ is
a parameter introduced to stress a quasiclassical
nature of the large $N$ limit.
The partition function is defined as an integral
\beq\label{partition}
Z_N = \int [D \Phi ] e^{\frac{1}{\hbar}\mbox{tr}\, W(\Phi )}
\eeq
over the matrices with the integration measure
$[D \Phi ]$ to be specified below in this section.

We consider two ensembles of random matrices $\Phi$ with
complex eigenvalues: ensemble $\complex$ of {\it general
complex matrices} (with no restrictions on the entries
except for $\det \Phi \neq 0$) and ensemble $\normal$
of {\it normal matrices} (such that $[\Phi , \Phi ^{\dag}]=0$).

\subsection{Integration measure}

The integration measure has the most simple form
for the ensemble of general complex matrices:
$$
[D \Phi ]=\prod_{i,j =1}^{N}
d(\Re \, \Phi_{ij})\, d (\Im \, \Phi_{ij}).
$$
This measure is additively invariant and multiplicatively
covariant, i.e.
for any fixed (non-degenerate) matrix $A\in \complex$ we have
the properties
$[D(\Phi +A)] =[D\Phi ]$ and
$[D(\Phi A)] =[D(A\Phi )]=|\det A|^{2N} [D\Phi ]$.
It is also clear that the measure is invariant under
transformations of the form $\Phi \to U^{\dag}\Phi U$ with a unitary
matrix $U$.

The measure for $\normal$ is induced by the
standard flat metric in $\complex$,
$
||\delta \Phi ||^2 =\mbox{tr} \,
(\delta \Phi \delta \Phi^{\dag}) =\sum_{ij} |\delta \Phi_{ij}|^2
$
via the embedding
$\normal \subset \complex$.
Here $\normal$ is regarded as a hypersurface in $\complex$ defined by
the quadratic relations $\Phi \Phi^{\dag}=\Phi^{\dag}\Phi$.
As usual in the theory of random matrices,
one would like to integrate out
the ``angular'' variables and to express the integration measure
through the eigenvalues only.

\paragraph{The measure for $\normal$ through
eigenvalues \cite{Mehta,normal}.}
We derive the explicit representation of the measure in terms
of eigenvalues in three steps:
\begin{itemize}
\item[1.]
Introduce coordinates in $\normal \subset \complex$.
\item[2.]
Compute the inherited metric on $\normal$ in these
coordinates: $||\delta \Phi ||^2 =g_{\alpha \beta}
d\xi^{\alpha}d\xi^{\beta}$.
\item[3.]
Compute the volume element
$[D\Phi ]=\sqrt{|\det g_{\alpha \beta}|}
\prod_{\alpha} d\xi^{\alpha}$.
\end{itemize}

\noindent
{\it Step 1: Coordinates in $\normal$}.
For any matrix $\Phi$, the matrices
$H_1 =\frac{1}{2}(\Phi +\Phi^{\dag})$,
$H_2 =\frac{1}{2i}(\Phi -\Phi^{\dag})$
are Hermitian.
The condition
$[\Phi , \Phi^{\dag}]=0$ is equivalent to
$[H_1 , H_2]=0$.
Thus $H_{1,2}$ can be simultaneously diagonalized
by a unitary matrix $U$:
$$
\begin{array}{l}
H_1 =UXU^{\dag}\,,
\quad X=\mbox{diag}\, \{ x_1 , \ldots , x_N\}\\
H_2 =UYU^{\dag}\,,
\quad \, \, Y=\mbox{diag}\, \{ y_1 , \ldots , y_N\}\,.
\end{array}
$$
Introduce the diagonal matrices
$Z=X+iY$, $\bar Z=X-iY$ with diagonal elements
$z_j =x_j +iy_j$ and $\bar z_j =x_j - iy_j$ respectively.
Note that $z_j$ are eigenvalues of $\Phi$.
Therefore, any $\Phi \in \normal$
can be represented as
$
\Phi =UZU^{\dag}
$,
where $U$ is a unitary matrix and $Z$ is the diagonal
matrix with eigenvalues of $\Phi$ on the diagonal.
The matrix $U$ is defined up to multiplication by a diagonal
unitary matrix from the right:
$U \rightarrow U \, U_{{\rm diag}}$.
The dimension of $\normal$ is thus
$$
\mbox{dim}\, (\normal )=
\mbox{dim}\, (\unitary ) -
\mbox{dim}\, (\unitary_{{\rm diag}} )+
\mbox{dim}\, (\complex_{{\rm diag}} )=
N^2 -N + 2N \, =\, N^2 +N
$$
(here $\unitary \subset \complex$ is the submanifold
of unitary matrices).

\noindent
{\it Step 2: The induced metric}.
Since $\Phi =UZU^{\dag}$, the variation is
$\delta \Phi =U(\delta u \cdot Z + \delta Z + Z \cdot \delta
u^{\dag})U^{\dag}$,
where
$\delta u^{\dag}=U\delta U^{\dag}=-\delta u^{\dag}$.
Therefore,
$$
||\delta \Phi ||^2 =\, \mbox{tr}\,
(\delta \Phi \delta \Phi^{\dag})
=\, \mbox{tr}\, (\delta Z \delta \bar Z) +2\, \mbox{tr}\,
(\delta u Z \delta u \bar Z \! -\! (\delta u)^2 Z \bar Z)
$$
$$
=\sum_{j=1}^{N}|\delta z_j |^2 + 2 \sum_{j<k}^{N}
|z_j -z_k |^2 \, |\delta u_{jk}|^2.
$$
(Note that $\delta u_{jj}$ do not enter.)

\noindent
{\it Step 3. The volume element}.
We see that the metric $g_{\alpha \beta}$ is diagonal
in the coordinates
$\Re (\delta z_j)$,
$\Im (\delta z_j)$,
$\Re (\delta u_{jk})$,
$\Im (\delta u_{jk})$ with $1\leq j<k\leq N$, so the determinant
of the diagonal matrix $g_{\alpha \beta}$ is
easily calculated to be
$|\det g_{\alpha \beta}|=2^{N^2 -N}
\prod_{j<k}^{N} |z_i -z_k |^4$. Therefore,
\beq\label{measure1}
[D \Phi ]\propto [D U]' \, |\Delta_N (z_1 , \ldots , z_N )|^2
\prod_{j=1}^{N} d^2 z_j\,,
\eeq
where $d^2 z\equiv dx dy$ is the
flat measure in the complex plane,
$[DU]' = [DU]/ [DU_{{\rm diag}}]$
is the invariant measure on $\unitary / \unitary _{{\rm diag}}$,
and $\Delta_N$ is the Vandermonde determinant:
\beq\label{Vandermonde}
\Delta_N (z_1 , \ldots , z_N)=\prod_{j>k}^{N} (z_j \! -\! z_k )
=\det_{N\times N} (z_{j}^{k-1}).
\eeq

\paragraph{The measure for $\complex$ through eigenvalues.}
A complex matrix $\Phi$ with eigenvalues $z_1 , \ldots , z_N$
can be decomposed as
$
\Phi =U(Z+R)U^{\dag}
$,
where $Z=\mbox{diag}\, \{ z_1 , \ldots , z_N\}$ is
diagonal, $U$ is unitary, and $R$ is strictly upper triangular, i.e.,
$R_{ij}=0$ if $i\geq j$. These matrices are defined up to a
``gauge transformation":
$U\to U\, U_{{\rm diag}}$,
$R\to U^{\dag}_{{\rm diag}} \, R \, U_{{\rm diag}}$.
It is not so easy to see that the measure factorizes.
This requires some work, of which the key
step is a specific ordering of the independent variables.
The final result is:
\beq\label{measure3}
[D \Phi ] \propto
[DU]' \, \left ( \prod_{k<l} d^2 R_{kl}\right )
|\Delta_{N}(z_i)|^2 \prod_{j=1}^{N} d^2 z_j\,.
\eeq
The details can be found in the Mehta book \cite{Mehta}.

\subsection{Potentials}

For the ensemble $\normal$ the
``angular variables" (parameters of the unitary matrix $U$)
always decouple after taking the trace
$\mbox{tr}\, W(\Phi )=\sum_j W(z_j)$,
so the potential $W$ can be a function
of $\Phi$, $\Phi^{\dag}$ of a general form
$W (\Phi ) =\sum a_{nm} \Phi^n (\Phi^{\dag})^m$.
The partition function reads
\beq\label{Z}
Z_N =\int |\Delta_N (z_i)|^{2}
\prod _{j=1}^{N}
e^{\frac{1}{\hbar}W(z_j)} d^2 z_j,
\eeq
where we ignore a possible
$N$-dependent normalization factor.
From now on this formula is taken as the definition of the
partition function.

The choice of the potential for the
ensemble $\complex$ is more restricted.
For a general potential,
the matrix $U$ in
$\Phi =U(Z+R)U^{\dag}$
still decouples but $R$ does not.
An important class of potentials when $R$ decouples nevertheless
is $W(\Phi )=-\Phi \Phi^{\dag} +V(\Phi ) +
\bar V(\Phi ^{\dag})$, where
$V(z)$ is an analytic function of $z$ in some
domain containing the origin and
$\bar V(z)=\overline{V(\bar z)}$.
In terms of the eigenvalues,
\beq\label{quasiharm}
W(z )=-|z|^2 +V(z ) +
\overline{V(z)}.
\eeq
In what follows, we call such a potential {\it quasiharmonic}.
In this case,
$
\mbox{tr}\, (\Phi \Phi^{\dag})=
\mbox{tr}\, (Z\bar Z)+
\mbox{tr}\, (R R^{\dag})$,
$\mbox{tr}\, (\Phi ^n)=
\mbox{tr}\, (Z+R)^n =\mbox{tr}\, Z^n$,
and so
\beq\label{measure4}
\int_{\complex} [D\Phi ] e^{\frac{1}{\hbar}\mbox{tr}\, W(\Phi )}=
C_N
\int |\Delta (z_i)|^2 \prod_k e^{\frac{1}{\hbar}W(z_k)}d^2 z_k,
\eeq
where $C_N$ is an $N$-dependent normalization factor
proportional to the gaussian integral
$\int [D R] e^{-\frac{1}{\hbar}\mbox{tr}\, (RR^{\dag})}$.

As an example, let us consider the quadratic potential:
$
W(z)=- |z|^2 +2\Re \, (t_1 z +t_2 z^2 )
$.
The ensemble $\complex$ with this potential
is known as {\it the Ginibre-Girko ensemble} \cite{Ginibre,Girko}.
In this case the partition function (\ref{Z})
can be calculated exactly \cite{FGIL}:
\beq\label{Ginibre-Girko}
Z_N =
 Z_{N}^{(0)}
(1 -4|t_2|^2 )^{-N^2 /2}
\exp \left (\frac{N}{\hbar}\,
\frac{t_{1}^{2}\bar t_2
+\bar t_{1}^{2}t_2 + |t_1|^2}{1 -4|t_2|^2}
\right ),
\eeq
where
$
\displaystyle{Z_{N}^{(0)}
=\hbar ^{(N^2 +N)/2} \pi^N \prod_{k=1}^{N}k!}
$
is the partition function of the model with
$W=-|z|^2$.

Note that for quasiharmonic potentials the integral (\ref{Z})
diverges unless $V$ is quadratic or logarithmic with suitable
coefficients. The simplest way to give sense to the integral
when it diverges at infinity is to introduce a cut-off, i.e.,
to integrate over a big but finite disk of radius $R_0$ centered
at the origin. Just for technical simplicity we assume that
a) $V$ is a holomorphic function everywhere inside this disk,
b) $W$ has a maximum at the origin with $W(0)=0$
and no other critical critical
points inside the disk, c) At $|z|=R_0$ the potential $W$ is bounded
from above by a constant $B<0$.
The large $N$ expansion is then well-defined.
For details and rigorous proofs see
\cite{rigorous1,rigorous2,rigorous3}.

\subsection{The Dyson gas picture}

The statistics of eigenvalues
appears to be mathematically equivalent to
some important models of classical statistical mechanics,
with the eigenvalues being represented as charged particles
in the plane interacting via 2D Coulomb (logarithmic) potential.
This interpretation, first suggested by Dyson \cite{Dyson}
for the unitary, symplectic and orthogonal matrix ensembles,
relies on
rewriting $|\Delta_N (z_i)|^{2}$ as
$\exp \Bigl ( \sum_{i\neq j}\log |z_i \! - \! z_j | \Bigr )$.
Clearly, the integral (\ref{Z}) looks then
exactly as the partition function of the
2D Coulomb plasma (often called the Dyson gas)
in the external field:
\beq\label{ZE}
Z_N =\int e^{-\beta E(z_1 , \ldots , z_N)}
\prod d^2 z_j\,,
\eeq
where
\beq\label{energy}
E=-\sum_{i<j}\log |z_i -z_j |^2
-\frac{1}{\beta \hbar}\sum_j W(z_j).
\eeq
Here $\beta$ plays the role of inverse temperature
(in (\ref{Z}) $\beta =1$).
The first sum is the Coulomb interaction energy, the
second one is the energy due to the external field.
For the Hermitian and unitary ensembles
the charges are confined to lines of dimension 1
(the real line or the unit circle)
but still interact as 2D Coulomb charges. So, the Dyson gas picture
for ensembles of matrices with general complex eigenvalues
distributed on the plane
looks even more natural. It becomes
especially helpful in the large $N$ limit, where it allows
one to apply thermodynamical arguments.

The Dyson gas picture prompts to consider more general
ensembles with arbitrary values of $\beta$ (``$\beta$-ensembles"):
\beq\label{Zbeta}
Z_N =\int |\Delta_N (z_i)|^{2\beta}
\prod _{j=1}^{N}
e^{\frac{1}{\hbar}W(z_j)} d^2 z_j\,.
\eeq
In general they can not be defined through
matrix integrals. As we shall see, the leading large $N$ contribution
has a simple regular dependence on $\beta$. However, the sub-leading
corrections may depend on $\beta$ in a rather non-trivial way.

\section{Exact relations at finite $N$}

Here we present some general exact relations
for correlation functions valid for any values of $N$
and $\beta$.

\subsection{Correlation functions: general relations}

The main objects of interest are
correlation functions, i.e., mean values of
functions of matrices.
We shall consider functions that depend on
eigenvalues only -- for example, traces $\mbox{tr}\, f(\Phi )=\sum_i f(z_i)$.
Here, $f(\Phi )=f(\Phi, \Phi^{\dag})$
is any function of $\Phi$, $\Phi^{\dag}$ which is regarded as
the function $f(z_i)=f(z_i, \bar z_i)$
of the complex argument $z_i$ (and $\bar z_i$).
(For the abuse of notation, in case of arbitrary $\beta$
we write $\mbox{tr}\, f := \sum_i f(z_i)$ although
a matrix realization may be not available.)
Typical correlation functions which we are
going to study are mean values of products of traces:
$\lbracket \mbox{tr}\, f(\Phi )\rbracket$,
$\lbracket \mbox{tr}\, f_1(\Phi )
\, \mbox{tr}\, f_2(\Phi ) \rbracket $
and so on.
Clearly, they
are represented as integrals over eigenvalues. For instance,
$$
\lbracket \mbox{tr}\, f(\Phi )\rbracket =\frac{1}{Z_N}
\int |\Delta_N (z_i)|^{2\beta}
\left ( \sum_{l=1}^{N}f(z_l)\right ) \prod_{j=1}^{N}
e^{\frac{1}{\hbar}W(z_j)}d^2 z_j\,.
$$
A particularly important example is the density function
defined as
\beq\label{density1}
\rho (z)=\hbar \sum_j \delta ^{(2)} (z-z_j),
\eeq
where $\delta ^{(2)}(z)$ is the 2D $\delta$-function.
Note that in our units $\hbar$ has dimension of $[\mbox{length}]^2$
and $\rho (z)$ is dimensionless.
As it immediately follows from the
definition, any correlator of traces is expressed through
correlators of $\rho$:
\beq\label{trf}
\lbracket \mbox{tr}\, f_1  \, \ldots \,
\mbox{tr}\, f_n \rbracket
=\hbar^{-n}\int \lbracket \rho (z_1) \ldots \rho (z_n)\rbracket
f_1 (z_1 ) \ldots f_n (z_n ) \prod_{j=1}^{n}d^2 z_j\,.
\eeq
Instead of correlations of density it is often convenient
to consider correlations of the field
\beq\label{potential1}
\varphi (z)=-\beta \hbar \sum_j \log |z-z_j |^2
\eeq
from which the correlations of density can be found
by means of the relation
\beq\label{rhophi}
4\pi \beta \rho (z)=-\Delta \varphi (z),
\eeq
where $\Delta =4\p \bar \p$ is the Laplace operator.
Clearly, $\varphi$ is the 2D Coulomb potential
created by the eigenvalues (charges).

Handling with multi-point correlation functions,
it is customary to pass to their {\it connected parts}.
For example, in the case of 2-point functions, the
connected correlation function is defined as
$$
\lbracket \rho (z_1 )\rho (z_2)\rbracket _{c}\equiv
\lbracket \rho (z_1 )\rho (z_2)\rbracket -
\lbracket \rho (z_1) \rbracket
\lbracket \rho (z_2) \rbracket .
$$
The connected multi-trace correlators are expressed
through the connected density correlators by the same
formula (\ref{trf}) with $\lbracket \rho (z_1)\ldots
\rho (z_n) \rbracket_c$ in the r.h.s.
The connected part of the
$(n+1)$-point density correlation function is given by the
linear response of the $n$-point one to a small variation
of the potential.
More precisely,
the following variational formulas
hold true:
\beq\label{var}
\lbracket \rho (z) \rbracket =\hbar^2 \,
\frac{\delta \log Z_N}{\delta W(z)}\,,
\;\;\;\;
\lbracket \rho (z_1 )\rho (z_2)\rbracket _{c}=\hbar^2 \,
\frac{\delta \lbracket \rho (z_1)\rbracket }{\delta W(z_2)}=\hbar^4
\frac{\delta^2 \log Z_N}{\delta W(z_1) \delta W(z_2)}.
\eeq
Connected multi-point correlators are higher
variational derivatives of $\log Z_N$.
These formulas follow from the fact that
variation of the partition function over a general
potential $W$ inserts $\sum_i \delta ^{(2)}(z-z_i)$ into
the integral. Basically, they are linear response relations used
in the Coulomb gas theory \cite{Forrester}.

\subsection{Loop equations}

The standard source of exact relations
for correlation functions is
the formal identity
\beq\label{id0}
\sum_i \int \frac{\p}{\p z_i} \left (
\epsilon (\{z_j\})\, e^{-\beta E}\right ) \prod_j d^2 z_j =0,
\eeq
where $\epsilon (\{z_j\})$ is any function of coordinates $z_j$
bounded at infinity and
$E$ is given by (\ref{energy}).
Introducing, if necessary, a cutoff
at infinity one sees that
the 2D integral over $z_i$
can be transformed, by virtue of the Green's theorem,
into a contour integral around infinity and so it does
vanish. Being expressed in terms of correlation functions
of local fields (such as $\rho$ or $\varphi$), this identity
yields, with a suitable choice of $\epsilon$, certain exact
relations between them. For historical reasons,
they are referred to as {\it loop equations}.

Let us take
\beq\label{g1}
\epsilon (z_i )=\frac{X(\{z_l\})}{z-z_i},
\eeq
where $X(\{z_l\})$ is any symmetric function of
$z_1, z_2, \ldots , z_N$, then the identity reads
$$
\sum_i \int \left [ \left (
\frac{-\beta \, \p_{z_i}E}{z-z_i}+
\frac{1}{(z-z_i)^2} \right )X + \frac{\p_{z_i}X}{z-z_i}
\right ] e^{-\beta E} \prod_j d^2 z_j =0.
$$
The singularity
at the point $z$ does not destroy the identity
since its contribution is proportional to
the vanishing integral
$\oint d\bar z_i /(z_i -z)$ over a small contour
encircling $z$.
Plugging
$\p_{z_i}E$
in the first term and
using the bracket notation for the mean value, we have:
$$
\lbracket
\left (\frac{1}{\hbar}\sum_i \frac{\p W(z_i)}{z-z_i}
+\beta \sum_{i\neq j} \frac{1}{(z-z_i)(z_i -z_j)}
+\sum_i \frac{1}{(z-z_i)^2}\right )X +\sum_i \frac{\p_{z_i}X}{z-z_i}
\rbracket =0.
$$
The second sum can be transformed by means of the simple
algebraic identity
$$
\sum_{i,j}\frac{1}{(z-z_i)(z-z_j)}
=\sum_{i\neq j}\frac{2}{(z-z_i)(z_i -z_j)}+
\sum_i \frac{1}{(z-z_i)^2}\,.
$$
Finally, we arrive at the relation
\beq\label{g3}
\lbracket L(z)X + \sum_i
\frac{\p_{z_i}X}{z-z_i} \rbracket =0,
\eeq
where we have introduced the special notation
\beq\label{g3a}
L(z)=\frac{1}{\hbar}\sum_i \frac{\p W(z_i)}{z-z_i}
+\frac{\beta}{2} \left (\sum_i \frac{1}{z-z_i}\right )^2
+\left (1-\frac{\beta}{2}\right )\sum_i  \frac{1}{(z-z_i)^2}\,,
\eeq
or, in terms of the fields $\rho$, $\varphi$ (\ref{density1}),
(\ref{potential1}),
\beq\label{g4}
2\beta \hbar^2 L(z)=2\beta \int \frac{\p W(\xi )
\rho (\xi )d^2 \xi}{z-\xi}
\, +\, (\p \varphi (z))^2 \, +\, (2\! -\! \beta )
\hbar \p^2 \varphi (z).
\eeq
This quantity plays an important role. It is to be compared
with the stress energy tensor in 2D CFT.

We call (\ref{g3}) the generating loop equation. It
generates an infinite
hierarchy of identities obeyed by correlation functions. The simplest
one is obtained at $X\equiv 1$: $\lbracket L(z)\rbracket =0$.
It reads:
\beq\label{loopeq}
\frac{1}{2\pi}\int \frac{\p W(\zeta ) \lbracket
\Delta \varphi (\zeta )\rbracket}{\zeta -z} \, d^2 \zeta +
\lbracket
(\p \varphi (z))^2
\rbracket +(2\! -\! \beta )\hbar \, \lbracket
\p^2 \varphi (z)\rbracket =0.
\eeq
This identity gives an exact relation between one- and two-point
correlation functions because
the mean value
$\lbracket (\p \varphi (z))^2 \rbracket$
can be reproduced from the two-point correlation function
$\lbracket \p \varphi (z)\, \p \varphi (z')\rbracket$
by averaging over all possible directions of approaching $z' \to z$:
\beq\label{limint}
\lbracket (\p \varphi (z))^2 \rbracket =
\lim _{\varepsilon \to 0}\,
\frac{1}{2\pi}\int_{0}^{2\pi}
\lbracket \p \varphi (z)\, \p
\varphi (z\! +\! \varepsilon e^{i\theta})\rbracket  d\theta \,.
\eeq
Another interesting choice is $X=\varphi (\zeta )$, where
$\zeta \neq z$, then
$
\p_{z_i}X=\beta \hbar /(\zeta -z_i)
$
and (\ref{g3}) yields
\beq\label{g5}
\lbracket L(z)\varphi (\zeta ) +\,
\frac{\p \varphi (z)-\p \varphi (\zeta )}{z-\zeta}
\rbracket =0.
\eeq
Acting by $\p_{\bar \zeta}$, we get
$
\lbracket L(z)\bar \p \varphi (\zeta )\rbracket =
\pi \beta  \!
\lbracket \rho (\zeta )\rbracket /(\zeta-z)
$.
Further, acting by $\p _{\zeta}$ to both sides,
we obtain the relation
\beq\label{g7}
\lbracket L(z) \rho (\zeta )\rbracket =
\frac{\lbracket \rho (\zeta )\rbracket}{(z-\zeta )^2}+
\frac{\p_{\zeta}\lbracket \rho (\zeta )\rbracket}{z-\zeta}\,.
\eeq
A similar but longer calculation gives a
generalization of this identity for
$m$-point functions:
\beq\label{g8}
\lbracket L(z) \rho (\zeta _1 )\ldots \rho (\zeta _m )\rbracket =
\sum_{i=1}^{m}\left ( \frac{1}{(z-\zeta _i )^2}+
\frac{1}{z-\zeta _i}\, \p_{\zeta _i}\right )
\lbracket  \rho (\zeta _1 )\ldots \rho (\zeta _m )\rbracket .
\eeq
It has the form of the conformal Ward identity for
primary field of conformal dimensions $1$, with $L(z)$
playing the role of the
(holomorphic component of) the stress energy tensor in CFT.

\section{Large $N$ limit}

Starting from this section, we study the
large $N$ limit
\beq\label{large}
N \to \infty \,, \quad \hbar \to 0 \,,
\quad \hbar N =t \quad \mbox{finite}.
\eeq
We shall see that in this limit
meaningful analytic and algebro-geometric structures
emerge, as well as important applications in physics.

\subsection{Solution to the loop equation in the leading order}

It is instructive to think about the large $N$ limit
under consideration
in terms of the Dyson gas picture.
Then the limit we are interested in corresponds to
a very low temperature of the gas, when fluctuations
around equilibrium positions of the charges are
negligible. The main contribution to the partition function
then comes from a configuration,
where the charges are ``frozen" at their equilibrium
positions. It is also important that
the temperature tends to zero simultaneously with increasing the
number of charges, so the plasma can be regarded as a
continuous fluid at static equilibrium.
Mathematically, all this means that the integral is
evaluated by the saddle point method, with only the
leading contribution being taken into account.
As $\hbar \to 0$, correlation functions
take their ``classical" values
$\lbracket \varphi (z) \rbracket = \vphcl (z)$,
$\lbracket \rho (z) \rbracket = \rhocl (z)$,
and multi-point functions factorize in the
leading order:
$\lbracket \p \varphi (z) \,
\p \varphi (z')\rbracket =
\p \vphcl (z) \p\vphcl (z')$, etc.
Then the loop equation (\ref{loopeq})
becomes a closed relation for $\vphcl$:
\beq\label{loopeq2}
\frac{1}{2\pi} \int \frac{\p W( \zeta )\Delta
\vphcl (\zeta )}{\zeta -z}\, d^2 \zeta \,
+\, \Bigl ( \p \vphcl (z)\Bigr )^2 \, = \, 0,
\eeq
where we have ignored the last term because it has
a higher order in $\hbar$.
Applying $\bar \p$ to the both terms, we get:
$
-\p W(z) \Delta \vphcl (z)+
\p \vphcl (z) \Delta \vphcl (z)=0
$.
Since $\Delta \vphcl (z) =-4\pi \beta \rhocl (z)$
(see (\ref{rhophi})),
we obtain
\beq\label{large1}
\rhocl (z) \, \left [ \p \vphcl (z) -\p W(z)\right ]=0.
\eeq
This equation should be solved with the
additional constraints
$\int \rhocl (z) d^2 z =t$ (normalization) and
$\rhocl (z)\geq 0$.
The equation tells us that
$\mbox{either} \quad \p \vphcl (z)=\p W(z)$ or
$\rhocl (z)=0$.
Applying $\bar \p$ to the former,
we get $\Delta \vphcl (z)=\Delta W(z)$.
This gives the solution for $\rhocl$:
\beq\label{large2}
\rhocl (z)=-\, \frac{\Delta W(z)}{4\pi \beta}
\quad \quad \mbox{in the bulk.}
\eeq
Here, ``in the bulk" means ``in the region where
$\rhocl > 0$".
The physical meaning of the equation
$\p \vphcl (z) =\p W(z)$
is clear. It is just the condition that
the charges are in equilibrium (the saddle point for the
integral).
Indeed, the equation states that the total force
experienced by a charge at any point $z$ where
$\rhocl \neq 0$
is zero, i.e., the interaction with the other charges,
$\p \vphcl (z)$, is compensated by the
force $\p W(z)$ due to the external field.

\subsection{Support of eigenvalues}

Let us assume that
\beq\label{sigma}
\sigma (z):=-\frac{1}{4\pi}\Delta W(z) >0.
\eeq
For quasiharmonic potentials, $\sigma (z)=1/\pi$.
If, according to (\ref{large2}),
$\rhocl =\sigma /\beta$ everywhere, the normalization
condition for $\rhocl$ in general can not be satisfied.
So we conclude that
$\rhocl = \sigma /\beta$ in a compact bounded domain
(or domains) only, and outside this domain one should
switch to the other solution of (\ref{large1}),
$\rhocl =0$.
The domain ${\sf D}$ where $\rhocl >0$ is called
{\it support of eigenvalues} or droplet of eigenvalues.
In general, it may
consist of several disconnected components.
The complement to the support of eigenvalues,
${\sf D^c} = \CC \setminus {\sf D}$, is an unbounded
domain in the complex plane.
For quasiharmonic potentials, the result is especially
simple: $\rhocl$ is constant in ${\sf D}$ and
$0$ in ${\sf D^c}$.

To find the shape of ${\sf D}$ is a much more
serious problem. It appears to be equivalent to the inverse
potential problem in 2D.
The shape of ${\sf D}$ is determined by the condition
$\p \vphcl (z) =\p W(z)$ (valid at all points $z$
inside ${\sf D}$)
and by the normalization condition. One can write
them in the form
\beq\label{large3a}
\left \{
\begin{array}{l}
\displaystyle{
\frac{1}{4\pi} \int_{{\sf D}}
\frac{\Delta W(\zeta )d^2 \zeta}{z-\zeta} \, = \,
\p W(z)} \quad \quad \mbox{for all $z \in {\sf D}$}
\\  \\
\displaystyle{
\int_{{\sf D}}
\sigma (\zeta ) d^2 \zeta =\beta t \,.}
\end{array}
\right.
\eeq
The integral
over ${\sf D}$ in the first equation
can be transformed to a contour integral
by means of the Cauchy formula.
As a result, the first equation reads:
\beq\label{large3}
\oint_{\p {\sf D}} \frac{\p W(\zeta )d\zeta}{z-\zeta}
=0
\quad \quad \mbox{for all $z\in {\sf D}$.}
\eeq
This means that the domain ${\sf D}$
has the following property:
the function $\p W(z)$
on its boundary is the boundary value of an analytic
function in its complement ${\sf D^c}$.

The connection with the inverse potential problem
is most straightforward in
the quasiharmonic case, where
$\p W(z)=-\bar z +V'(z)$,
$\Delta W(z)=4\p\bar \p W(z)=-4$.
The normalization then means that the area of ${\sf D}$
is equal to $\beta \pi t$.
Assume that: i) $V(z)=\sum _k t_k z^k$ is regular in ${\sf D}$
(say a polynomial), ii) $0\in {\sf D}$
(it is always the case when $W$ has a
maximum at $0$), iii) ${\sf D}$ is connected.
Then  equation (\ref{large3})
acquires the form
$\displaystyle{
\frac{1}{2\pi i}\oint_{\p {\sf D}}
\frac{\bar \zeta d \zeta}{\zeta -z}=V'(z)}$
for $z \in {\sf D}$.
Expanding it near $z=0$, we get:
\beq\label{harmmom}
t_k =\frac{1}{2\pi i k}
\oint_{\p {\sf D}} \bar \zeta \zeta ^{-k} d\zeta =
-\frac{1}{\pi k} \int_{{\sf D}} \zeta ^{-k}d^2 \zeta \,.
\eeq
We see that the ``coupling constants" $t_k$
are {\it harmonic moments}
of ${\sf D^c}=\CC \setminus {\sf D}$
and the area of ${\sf D}$ is $\pi \beta t$.
It is the subject of the inverse potential problem
to reconstruct the domain from its
area and harmonic moments.
In general, the solution is not unique.
But it is known that {\it locally}, i.e.,
for a small enough change $t \rightarrow t+\delta t$,
$t_k \, \rightarrow \, t_k +\delta t_k$
there is only one solution.

\begin{figure}[tb]
\epsfysize=5cm
\centerline{\epsfbox{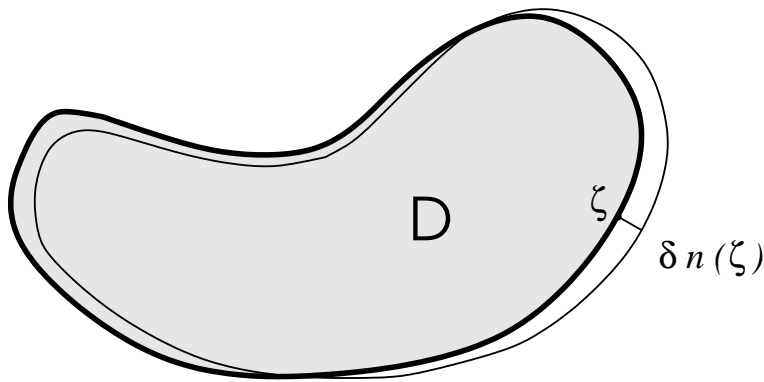}} \caption{\sl The
normal displacement of the boundary. } \label{fi:var}
\end{figure}

As an explicitly solvable example, consider the
Ginibre-Girko ensemble with the partition function
(\ref{Ginibre-Girko}). In this case the support of
eigenvalues is an ellipse with the half-axes
$$
a= \sqrt{ t \, \frac{1+2|t_2|}{1-2|t_2|}}, \quad \quad
b= \sqrt{ t \, \frac{1-2|t_2|}{1+2|t_2|}},
$$
centered at the point
$\displaystyle{
z_0 = \frac{2t_1 \bar t_2 +\bar t_1}{1-4|t_2|^2}}$, and with the
angle between the big axis and the real line being equal to
$\frac{1}{2} \mbox{arg} \, t_2$. For the model with the potential
$W=-|z|^2 + 2\Re (t_3 z^3 )$ the droplet of eigenvalues is bounded
by a hypotrochoid \cite{TBAZW05}.

\subsection{Small deformations of the support of eigenvalues}

Coming back to the general case,
let us examine how the shape of ${\sf D}$ changes
under a small change of the potential $W$
with $t=\hbar N$ fixed.
It is convenient to describe small deformations
${\sf D}\rightarrow \tilde {\sf D}$,
by the normal displacement $\delta n(\zeta )$
at a boundary point $\zeta$ (Fig.~\ref{fi:var}).
Consider a small variation of the potential
$W$ in the condition (\ref{large3}).
To take into account the deformation of the domain,
we write, for any fixed function $f$,
$$
\delta \left (
\oint_{\p {\sf D} }\!\! f(\zeta ) d \zeta \right ) =
\oint_{\p (\delta {\sf D})}\!\! f(\zeta ) d \zeta=2i\!\!
\int_{\delta {\sf D}}
\!\!\bar \p f(\zeta ) d^2 \zeta \approx 2i \!
\oint_{\p {\sf D}} \!\!\bar \p f(\zeta )\delta n(\zeta )
|d\zeta |
$$
(here $\delta {\sf D}=\tilde {\sf D} \setminus {\sf D}$)
and thus obtain from (\ref{large3}):
\beq\label{cor1}
\oint_{\p {\sf D}}\!\!
\frac{\p \, \delta W(\zeta ) d\zeta}{z-\zeta} \, + \,
\frac{i}{2}\oint_{\p {\sf D}}\!\!
\frac{\Delta W(\zeta ) \delta n(\zeta )}{z-\zeta}
\, |d\zeta | \, = \, 0.
\eeq
This integral equation for $\delta n (\zeta )$
can be solved in terms of the exterior
Dirichlet boundary value problem.
Given any smooth function $f(z)$,
let $f^H (z)$ be its
{\it harmonic continuation} from the boundary
of ${\sf D}$ to its exterior, i.e., a unique function such that
$\Delta f^H =0$ in ${\sf D^c}$
and regular at $\infty$, and
$f^H (z)=f(z)$ for all $z\in \p {\sf D}$.
Explicitly, a harmonic function
can be reconstructed from its boundary
value by means
of the formula
\beq\label{dirich}
f^H (z)=-\frac{1}{2\pi} \oint_{\p {\sf D}}
f(\xi )\p_n G(z, \xi) |d\xi | \,.
\eeq
(Here and below, $\p_n$
is the normal derivative at the
boundary, with the outward pointing
normal vector.)
The main ingredient of this formula
is the Green's function $G(z, \xi )$
of the domain
${\sf D^c}$ characterized by the properties
$\Delta_z G(z, \zeta )=2\pi \delta ^{(2)}(z-\zeta )$
in ${\sf D^c}$,
$G(z, \zeta )=0$ if $z$ or $\zeta$ $\in \p {\sf D}$.
As $\zeta \to z$, it has the logarithmic singularity
$G(z, \zeta )\to \log |z-\zeta |$.

Consider the integral
$\displaystyle{
\oint_{\p {\sf D}}\!\!
\frac{\p (\delta W^H (\xi )) d\xi}{z-\xi}}
$
which is obviously equal to $0$ for all $z$ inside ${\sf D}$,
subtract it from the first term in (\ref{cor1}) and
rewrite the latter as an integral over the
line element $|d\xi |$.
After this transformation
(\ref{cor1}) acquires the form
\beq\label{cor2}
\frac{1}{2\pi i}\oint_{\p {\sf D}}
\frac{R(\xi )}{z-\xi}
\, |d\xi | =0 \quad \quad
\mbox{for all $z\in {\sf D}$},
\eeq
where
\beq\label{Neumann}
R (z) =\Delta W (z) \delta n (z) +
\p_{n}^{-}\left (\delta W (z)\! -\! (\delta W)^H(z)\right )
\eeq
is a real-valued function on the boundary contour $\p {\sf D}$.
The superscript indicates that the derivative is
taken in the exterior of the boundary.
By properties of Cauchy integrals, it follows from
(\ref{cor2}) that
$ R(\xi )/\tau (\xi )$, where $\tau (\xi )=d\xi / |d\xi |$
is the unit tangential vector to the boundary curve,
is the boundary value of an analytic function $h(z)$ in ${\sf D^c}$
such that $h(\infty )=0$.
For $z\in {\sf D^c}$, this function is just given by the
integral in the l.h.s. of (\ref{cor2}).
Variation of
the normalization condition (the second equation
in (\ref{large3a})) yields,
in a similar manner:
\beq\label{cor3}
\oint_{\p {\sf D}}R(\xi )|d\xi |=0
\eeq
This relation implies that
the zero at $\infty$ is at least of the 2-nd order.

The following simple argument shows that an analytic function with
these properties must be identically zero.
Let $w(z)$ be the conformal map from ${\sf D^c}$
onto the unit disk such that $w(\infty )=0$
and the derivative at $\infty$ is real positive. By the
well known property of conformal maps we have
$$
\frac{dz}{|dz|}\, e^{i\,
\mbox{{\footnotesize arg}} w'(z)}=\frac{dw}{|dw|}
$$
along the boundary curve. Therefore,
$\tau (z)=i|w'(z)|\, w(z)/ w'(z)$
and we thus see that
$$
\frac{R(z)\, w'(z)}{i |w'(z)|w(z)}
$$
is the boundary value of the holomorphic function $h(z)$.
Since $w'(z)\neq 0$ in ${\sf D^c}$, the function
$
g(z)=h(z)\, w(z)/w'(z)
$
is holomorphic there with the
{\it purely imaginary} boundary value
$
\frac{R(z)}{i |w'(z)|}
$.
Then the real part of this function is harmonic
and bounded in ${\sf D^c}$ and is equal to $0$ on the
boundary. By uniqueness of a solution to the Dirichlet
boundary value problem, $\Re \, g(z)$ must be equal to $0$
identically. Therefore, $g(z)$ takes purely imaginary values
everywhere in ${\sf D^c}$ and so is a constant.
By virtute of
condition (\ref{cor3}) this constant must be $0$ which
means that $R(z) \equiv 0$.

Therefore, one obtains
the following result for the normal
displacement of the boundary
caused by a small change of the potential
$W \rightarrow W+\delta W$:
\beq\label{deltan}
\delta n (z)=\frac{\p_{n}^{-} (\delta W^H (z)\! - \!
\delta W(z))}{\Delta W(z)}\,.
\eeq

\subsection{From the support of eigenvalues to an
algebraic curve}

There is an interesting algebraic geometry behind the
large $N$ limit of matrix models.
For simplicity, here we
consider models with quasiharmonic potentials.

The boundary of the support of eigenvalues
is a closed curve in the plane without self-intersections.
If $V'(z)$ is a rational function, then
this curve
is a real section of a complex
algebraic curve of finite genus.
In fact, this curve encodes the
$1/N$ expansion of the model.
In the context of Hermitian 2-matrix model
such a curve was introduced and studied
in \cite{Staud,KM}.

To explain how the curve comes into play, we start from
the equation $\p \vphcl = \p W$, which
can be written in the form
$\bar z -V'(z)=G(z)$ for $z\in {\sf D}$, where
$\displaystyle{
G(z)=\frac{1}{\pi}\int_{{\sf D}}
\frac{d^2 \zeta}{z-\zeta}}
$.
Clearly, this function is analytic in ${\sf D^c}$.
At the same time,
$V'(z)$ is analytic in ${\sf D}$ and all its
singularities in
${\sf D^c}$ are poles.
Set
$
S(z)=V'(z)+G(z)
$.
Then
$
S(z)=\bar z
$
on the boundary of the support of eigenvalues.
So, $S(z)$ is the analytic continuation of $\bar z$
away from the boundary.
Assuming that poles of $V'$
are not too close to $\p {\sf D}$, $S(z)$ is
well-defined at least
in a piece of ${\sf D^c}$ adjacent to the boundary.
The complex conjugation yields $\overline{S(z)}=z$,
so the function $\bar S(z)=\overline{S(\bar z)}$ must
be inverse to the $S(z)$:
$
\bar S(S(z))=z
$
(``unitarity condition").
The function
$S(z)$ is called the {\it Schwarz function} \cite{Davis}.

Under our assumptions, $S(z)$ is an algebraic function,
i.e., it obeys a polynomial equation
$P(z, S(z))=0$ of the form
$$
P(z, S(z))=\sum_{n,l=1}^{d+1} a_{nl} z^n (S(z))^l =0,
$$
where $\overline{a_{ln}}= a_{nl}$ and
$d$ is the number of poles of $V'(z)$ (counted
with their multiplicities).
Here is a sketch of proof.
Consider the Riemann surface
$\mit\Sigma ={\sf D^c} \cup \p {\sf D} \cup ({\sf D^c})^*$
(the {\it Schottky double} of ${\sf D^c}$).
Here, $({\sf D^c})^*$
is another copy of ${\sf D^c}$, with the local coordinate $\bar z$,
attached to it along the boundary. On $\mit\Sigma$, there
exists an anti-holomorphic involution that interchanges
the two copies of ${\sf D^c}$ leaving the points of $\p {\sf D}$
fixed.
The functions $z$ and
$S(z)$ are analytically extendable to
$({\sf D^c})^*$ as $\overline{S(z)}$ and $\bar z$ respectively.
We have two meromorphic functions, each with $d+1$ poles,
on a closed Riemann surface. Therefore, they are connected by
a polynomial equation of degree $d+1$ in each variable.
Hermiticity of the coefficients follows from the unitarity condition.

The polynomial equation $P(z, \tilde z)=0$
defines a complex curve
$\mit\Gamma$
with anti-holomorphic involution
$(z, \tilde z)\mapsto
(\overline{\tilde z}, \bar z)$.
The real section is the set of points such that
$\tilde z = \bar z$.
It is the boundary of the support
of eigenvalues.

It is important to note that for models
with non-Gaussian weights (in particular,
with polynomial potentials of degree greater
than two) the curve has
a number of singular points, although the Riemann
surface $\mit\Sigma$ (the Schottky double) is smooth.
Generically, these are
{\it double points}, i.e., the points where the curve
crosses itself. In our case, a double point
is a point $z^{(d)}\in {\sf D^c}$
such that $S(z^{(d)})=\overline{z^{(d)}}$ but $z^{(d)}$
does not belong to the boundary of ${\sf D}$. Indeed,
this condition means that two different points of $\mit\Sigma$,
connected by the antiholomorphic involution,
are stuck together on the curve $\mit\Gamma$, which means
self-intersection.
The double points play the key role
in deriving the nonperturbative (instanton) corrections to
the large $N$ matrix models results (see \cite{nonperturbative}
for details).

\subsection{Free energy and correlation functions}

\paragraph{Free energy.}
The leading contribution to the free energy in the large $N$
limit is
$$
F_0 = \lim _{\hbar \to 0}\left ( \hbar^2 \log Z_N \right ).
$$
It is determined by the maximal value of the integrand
in (\ref{ZE}), i.e., by the extremum of the function
$E$. In the continuous approximation, one can represent it
as a functional of the density:
\beq\label{betaE1}
-\beta \hbar^2 E [\rho ]=\beta
\int\!\!\int \rho (z)\rho
(\zeta )\log |z-\zeta | d^2 z d^2 \zeta
\, +\, \int W(z)\rho (z) d^2 z
\eeq
and find its minimum with the constraint
$\int \rho \, d^2 z =t$. Introducing a Lagrange
multiplier $\lambda$, we get the equation
$$
2\beta \int \log |z-\zeta | \rho (\zeta ) d^2 \zeta + W(z) +\lambda =0
$$
which is solved, as expected,
by the function $\rho _0$ discussed in section 4.1.
Assuming that $W(0)=0$, we find $\lambda =\vphcl (0)$ and
\beq\label{free1}
F_0 = -\beta E[ \rhocl ]= - \frac{1}{\beta}\int_{{\sf D}}\!\! \int_{{\sf D}}
\sigma (z)\log \left | \frac{1}{z}\!-\! \frac{1}{\zeta}\right |
 \sigma (\zeta) d^2 z d^2 \zeta \,.
\eeq

\paragraph{Some results for correlation functions.}
Here are some results for the
correlation functions obtained by the variational
technique. (For details of the derivation
see \cite{WZnormal}.)
They are correct
at distances much larger than the mean distance
between the charges.
The leading contribution to the one-trace function
was already found in Section 4.1:
\beq\label{1trace}
\begin{array}{c}
\lbracket \mbox{tr} f(\Phi ) \rbracket \,\,=\,\,
\displaystyle{
\frac{1}{\hbar \beta}\int_{{\sf D}}
\sigma (z) f(z) \, d^2 z} +O(1).
\end{array}
\eeq
The connected two-trace function is:
\beq\label{2trace}
\lbracket \mbox{tr}\, f (\Phi )\, \mbox{tr}\, g(\Phi )\rbracket _{c}
=\frac{1}{4\pi \beta} \int_{{\sf D}} \nabla f \nabla g d^2 z-
\frac{1}{4\pi \beta} \oint_{\p {\sf D}} f \p_n g^H |dz|
+O(\hbar )
\eeq
(here $\nabla f \nabla g =\p_x f \p_x g + \p_y f \p_y g$).
In particular, for the connected
correlation functions of the fields
$\varphi (z)$
this formula gives:
\beq\label{corr2}
\!\!\! \lbracket \varphi (z)\varphi (z')\rbracket _{{\rm c}} \! =
\!\! \left \{
\begin{array}{l}
\displaystyle{\!\!- 2 \beta \hbar^2 \,
\log \frac{|z-z'|}{r} +O(\hbar^3) \quad \,\, \mbox{inside} \,\,  \DD }
\\ \\
\displaystyle{
\!\! 2 \beta \hbar^2 \! \left (
G(z,z') \! -\! G(z, \infty )\! -\! G(z' , \infty )
\! - \! \log \frac{|z\! -\! z'|}{r}
\right ) \! +\! O(\hbar^3) \,\,\, \mbox{outside} \,\, \DD }
\end{array}
\right.
\eeq
where $G$ is the Green's function of the Dirichlet
boundary value problem and
\beq\label{confrad}
\displaystyle{r=\exp \Bigl [ \lim_{\xi \to \infty}(
\log |\xi | +G(\xi , \infty ))\Bigr ]}
\eeq
is the external conformal radius of the domain ${\sf D}$
(the Robin's constant).
This result is valid if $|z-z'|\gg \sqrt{\hbar}$.
The 2-trace functions are {\it universal}, i.e.,
they depend on the shape of the support of eigenvalues only and
do not depend on the potential $W$ explicitly.
They resemble the two-point functions
of the Hermitian 2-matrix model found in \cite{DKK}; they were
also obtained in \cite{AlJan} in the study of thermal fluctuations of a
confined 2D Coulomb gas.
The structure of the formulas indicates
that there are local correlations in the
bulk as well as strong long range correlations at the edge
of the support of eigenvalues.
(See \cite{Jancovici82} for a similar result in the context
of classical Coulomb systems).

\section{The matrix model as a growth problem}

\subsection{Growth of the support of eigenvalues}

When $N$ increases at a fixed potential $W$,
one may say that the support of eigenvalues grows.
More precisely, we are going to find how
the shape of the support of eigenvalues
changes under
$t \rightarrow t+\delta t$, where $t=N\hbar$,
if $W$ stays fixed.

The starting point is the same as for the variations
of the potential, and the
calculations are very similar as well.
Variation of the conditions (\ref{large3a})
yields
$$
\oint_{\p {\sf D}}\frac{\Delta W(\zeta )\delta n(\zeta )}{z-\zeta}\,
|d\zeta| =0 \;\;\;\mbox{for all $z\in {\sf D}$},
\;\;\;\;\;
\oint_{\p {\sf D}}\Delta W(\zeta )\delta n(\zeta ) \, |d\zeta |
=-4\pi \beta \delta t\,.
$$
The first equation means that
$\Delta W(z)\delta n (z)
\frac{|dz|}{dz}$
is the boundary value of an analytic function
$h(z)$ such that
$h(z)=-4\pi \beta \delta t/z +O(z^{-2})$
as $z \to \infty$.
The solution for the $\delta n (z)$ is again expressed
in terms of the Green's function in ${\sf D^c}$:
\beq\label{gr0}
\delta n(z)=-\, \frac{\beta \delta t}{2\pi \sigma (z)} \,
\p_n G(\infty , z).
\eeq
For quasiharmonic potentials (with $\sigma =1/\pi$),
the formula simplifies:
\beq\label{gr1}
\delta n(z)=-\frac{\beta}{2} \, \delta t\,
\p_n G(\infty , z).
\eeq
Identifying $t$ with time,
one can say
that the normal velocity of the boundary at any point $z$,
$V_n (z)= \delta n (z)/ \delta t$, is
proportional to gradient of the Green's function:
$V_n (z)\propto -\p_n G(\infty , z)$.

If the domain is connected,
$G$ can be expressed through the conformal
map $w(z)$ from ${\sf D^c}$ onto the exterior
of the unit circle:
\beq\label{Gw}
G(z_1 , z_2 )=\log \left |
\frac{w(z_1 )-w(z_2 )}{1-w(z_1 ) \overline{w(z_2 )}}
\right |.
\eeq
In particular,
$G(\infty , z) =-\log |w(z)|$.
As $|z|\to \infty$,
$w(z)=z/r + O(1)$, where $r$ is the Robin's constant
which enters eq. (\ref{corr2}).
It is easy to see that $\p_n \log |w(z)| = |w'(z)|$ on $\p {\sf D}$,
so one can rewrite the growth law (\ref{gr1}) as
$\delta n(z)=\frac{\beta}{2} \,
|w'(z)|\, \delta t$.

\subsection{Laplacian growth}

The growth law (\ref{gr1}) is common to many important problems
in physics.
The class of growth processes, in which
dynamics of a moving front (an interface) between two distinct phases
is driven by a harmonic scalar field
is known under the name {\it Laplacian growth}.
The most known examples are
viscous flows in the Hele-Shaw cell,
filtration processes in porous media,
electrodeposition and
solidification of undercooled liquids.
A comprehensive list of relevant papers
published prior to 1998  can be found in
\cite{list}.

\paragraph{Viscous flows in Hele-Shaw cell.}
To be specific, we shall speak about an interface between two
incompressible fluids with very different viscosities on the plane
(say, oil and water).
In practice, the 2D geometry is
realized in the Hele-Shaw cell -- a narrow gap between two parallel
glass plates. For a review, see \cite{RMP,book}.
The velocity field in a viscous fluid in the Hele-Shaw cell
is proportional to the gradient of pressure $p$ (Darcy's law):
$$
\vec V =-K \nabla p\,,
\;\;\;\;
K=\frac{b^2}{12 \mu}\,.
$$
The constant
$K$ is called the filtration coefficient, $\mu$ is viscosity and
$b$ is the size of the gap between the two plates.
Note that if $\mu \to 0$, then $\nabla p \to 0$, i.e.,
pressure in a fluid with negligibly small viscosity is
uniform. Incompressibility of the fluids ($\nabla \vec V =0$)
implies that the
pressure field is harmonic: $\Delta p =0$.
By continuity, the velocity of the interface
between the two fluids is proportional to the normal
derivative of the pressure field on the boundary:
$V_n =-K \p_n p$.

\begin{figure}[tb]
\epsfysize=5cm
\centerline{\epsfbox{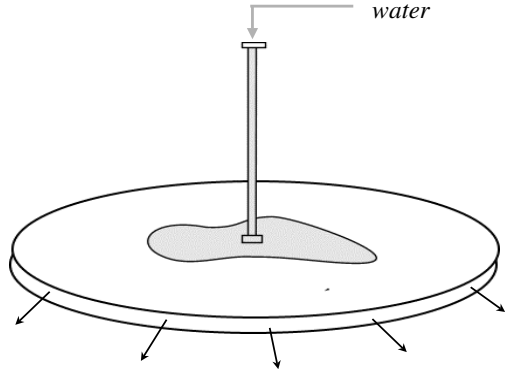}}
\caption{\sl
The Hele-Shaw cell.}
\label{fi:Hele-Shaw}
\end{figure}

To be definite, we assume that the Hele-Shaw cell
contains a bounded droplet
of water surrounded by an infinite ``sea" of oil
(another possible
experimental set-up is an air bubble surrounded
by oil or water).
Water is injected into the droplet while
oil is withdrawn at infinity
at a constant rate, as is shown
schematically in Fig.~\ref{fi:Hele-Shaw}.
The latter means that the pressure
field behaves as $p \propto -\log |z|$ at large distances.
We also assume that the interface
is a smooth closed curve $\Gamma$.
As it is mentioned above, one may set
$p=0$ inside the water droplet. However, pressure usually
has a jump across the interface, so $p$ in general
does not tend to zero if one approaches
the boundary from outside. This effect is due to
{\it surface tension}. It is hard to give realistic
estimates of the surface tension effect from fundamental principles, so
one often employs certain ad hoc assumptions.
The most popular one is to say that
the pressure jump is proportional to the local curvature, $\kappa$,
of the interface.

To summarize, the
mathematical setting of the Saffman-Taylor problem
is as follows:
\beq\label{lg1}
\left \{
\begin{array}{ll}
V_n =-\p_n p & \quad \mbox{on $\Gamma$}
\\
\Delta p =0 &\quad \mbox{in oil}
\\
p\to -\log |z| & \quad \mbox{in oil as $|z|\to \infty$}
\\
 p=0 & \quad \mbox{in water}
\\
p^{(+)}-p^{(-)}=-\nu \kappa & \quad \mbox{across $\Gamma$}
\end{array}
\right.
\eeq
Here $\nu$ is
the surface tension coefficient.
(The filtration coefficient is set to be $1$.)
The experimental evidence suggests that
when $\nu$ is small enough,
the dynamics becomes unstable. Any initial
domain develops an unstable fingering pattern.
The fingers split into new ones, and after a
long lapse of time the water droplet
attains a fractal-like structure.
This phenomenon is similar to
the formation of fractal patterns
in the diffusion-limited aggregation.

Comparing (\ref{gr1}) and (\ref{lg1}),
we identify ${\sf D}$ and ${\sf D^c}$ with
the domains occupied by water and oil
respectively, and conclude that the growth laws are
identical, with the pressure field being
given by the Green function:
$p(z) =G(\infty , z)$, and $p=0$ on the interface.
The latter means that supports of eigenvalues
grow according to (\ref{lg1})
with {\it zero surface tension}, i.e.,
with $\nu =0$ in (\ref{lg1}).
Neglecting the surface tension effects, one
obtains a good approximation unless the curvature
of the interface becomes large.
We see that the idealized
Laplacian growth problem, i.e., the one
with zero surface tension, is mathematically equivalent
to the growth of the support of eigenvalues
in ensembles of random matrices $\normal$
or $\complex$ and in general $\beta$-ensembles.

\paragraph{Finite-time singularities as critical points.}
As a matter of fact,
the Laplacian growth problem with zero surface tension
is ill-posed since an initially smooth interface
often becomes singular in the process of evolution,
and the solution blows up.
The role of surface tension is to inhibit
a limitless increase of the interface curvature.
In the absence of such a cutoff, the tip of
the most rapidly growing finger typically grows
to a singularity (a cusp).
In particular, a singularity necessarily occurs
for any initial interface that is the image of the
unit circle under a rational conformal map,
with the only exception of an ellipse.

An important fact is
that the cusp-like singularity occurs at a finite time $t=t_c$,
i.e., at a finite area of the droplet.
It can be shown that the conformal radius of the droplet $r$
(as well as some other geometric parameters),
as $t\to t_c$, exhibits a singular behavior
$$
r-r_c \propto (t_c -t)^{-\gamma}
$$
characterized by a critical exponent $\gamma$.
The generic singularity is the cusp $(2,3)$,
which in suitable local coordinates looks like
$y^2 =x^3$. In this case $\gamma =-\frac{1}{2}$.

A similar phenomenon has been known in the theory
of random matrices for quite a long time, and in fact it
was the key to their applications to 2D quantum gravity and string
theory. In the large $N$ limit, the random matrix models
have {\it critical points} -- the points
where the free energy is not analytic as a function
of a coupling constant. As we have
seen, the Laplacian growth time $t$
should be identified with a coupling constant
of the normal or complex matrix model.
In a vicinity of a critical point,
$$
F_0 \sim F_{0}^{{\rm reg}} \, + \alpha  (t_{c}-t)^{2-\gamma},
$$
where the critical index $\gamma$ (often
denoted by $\gamma_{{\rm str}}$ in applications to string theory)
depends on the type of the critical point.
Accordingly, the singularities show up in correlation
functions. Using the equivalence established above,
we can say that
the finite-time blow-up (a cusp-like singularity)
of the Laplacian growth with zero surface tension
is a critical point of the
normal and complex matrix models.
Remarkably, the evolution can be continued to the
post-critical regime as a dynamics of ``shock lines"
\cite{LTW09}.

\section*{Acknowledgments}

I am grateful to
O.Agam,
E.Bet\-tel\-heim,
I.Kos\-tov,
I.Kri\-che\-ver,
A.Mar\-sha\-kov,
M.Mi\-ne\-ev-\-Wein\-stein,
R.Teo\-do\-res\-cu
and P.Wieg\-mann
for collaboration.
The work was supported in part by RFBR grant 08-02-00287,
by grant for support of scientific schools
NSh-3035.2008.2 and by Federal Agency for Science and
Innovations of Russian Federation (contract 02.740.11.5029).

\end{document}